\documentclass[prb,twocolumn,showpacs,preprintnumbers,amsmath,amssymb,aps,10pt]{revtex4-1}

\usepackage{graphicx}
\usepackage{subfigure}
\usepackage{bm}
\usepackage[final=true]{hyperref}

\newcommand{\m}{\mathbf}
\newcommand{\be}{\begin{eqnarray}}
\newcommand{\ee}{\end{eqnarray}}
\newcommand{\nn}{\nonumber}

\begin{document}

\title{Mean-field approach for skyrmion lattice in hexagonal frustrated antiferromagnets}

\author{Oleg I. Utesov$^{1,2}$}
\email{utiosov@gmail.com}

\affiliation{$^1$National Research Center ``Kurchatov Institute'' B.P.\ Konstantinov Petersburg Nuclear Physics Institute, Gatchina 188300, Russia}
\affiliation{$^2$Department of Physics, Saint Petersburg State University, 198504 St.Petersburg, Russia}

\begin{abstract}

Simple mean-field approach for frustrated antiferromagnets on hexagonal lattices, aimed to describe the high-temperature part of the temperature-magnetic field phase diagram, is proposed. It is shown, that an interplay between modulation vector symmetry, Zeeman energy and magneto-dipolar interaction leads to stabilization of the triple-$Q$ skyrmion lattice in a certain region of the phase diagram. Corresponding analytical expressions for phase boundaries are derived. It is argued that the developed theory can be applied for the description of the high-temperature part of the phase diagram observed experimentally for Gd$_2$PdSi$_3$ compound.

\end{abstract}
\maketitle

\section{Introduction}
\label{Sintro}

Topological phases of matter is one of the hottest topics of contemporary solid state physics. In magnetism, individual skyrmions, their ordered arrays (skyrmion lattices, SkL), and various other topologically nontrivial structures are extensively studied (see, e.g., Refs.~\cite{bogdanov2020,gobel2021} and references therein).

In noncentrosymmetric magnets, skyrmions and SkL were predicted theoretically in seminal papers~\cite{bogdanov1989,bogdanov1994}. Experimental observation of SkL in the so-called \emph{A} phase of MnSi by means of elastic neutron scattering~\cite{muhlbauer2009} stimulates a plethora of further studies on this topic. Noteworthy, this interest is partially caused by promising technological applications (see, e.g., Refs.~\cite{fert2013, fert2017} and references therein).

Crucial skyrmion property relates to its nontrivial topology, which can be characterized by the topological charge~\cite{belavin1975metastable}
\begin{equation}\label{charge1}
  Q = \frac{1}{4 \pi} \int \m{n} \cdot \left[ \partial_x \m{n} \times \partial_y \m{n} \right] dx dy,
\end{equation}
where the $\m{n}=\m{s}/|\m{s}|$ is the unit vector along the local spin value, averaged over thermal and/or quantum fluctuations. Single skyrmion usually has $Q=\pm 1$. For SkL the natural measure of topological charge is its density $n_{sk}$, which, e.g., defines the topological contribution to the Hall resistivity, since $\rho^T \propto n_{sk}$~\cite{neubauer2009}. In noncentrosymmetric magnets the size of a magnetic unit cell is quite large, being of the order of $J/D \gg 1$ lattice parameters (here $J$ is some characteristic exchange energy, $D$ is a value of Dzyaloshinskii-Moriya interaction~\cite{dzyaloshinsky1958,moriya1960}), so $n_{sk}$ is somewhat suppressed. In contrast, in frustrated helimagnets modulation vectors are usually not small, which results in typical size of magnetic unit cell of the order of several nanometers (see Ref.~\cite{kurumaji2019Rev}). This leads to large $n_{sk}$ and the giant topological Hall effect, which were observed for the first time in Ref.~\cite{kurumaji2019SkL} in Gd$_2$PdSi$_3$ compound.

\begin{figure}
  \centering
  \hfill
  \includegraphics[width=4cm]{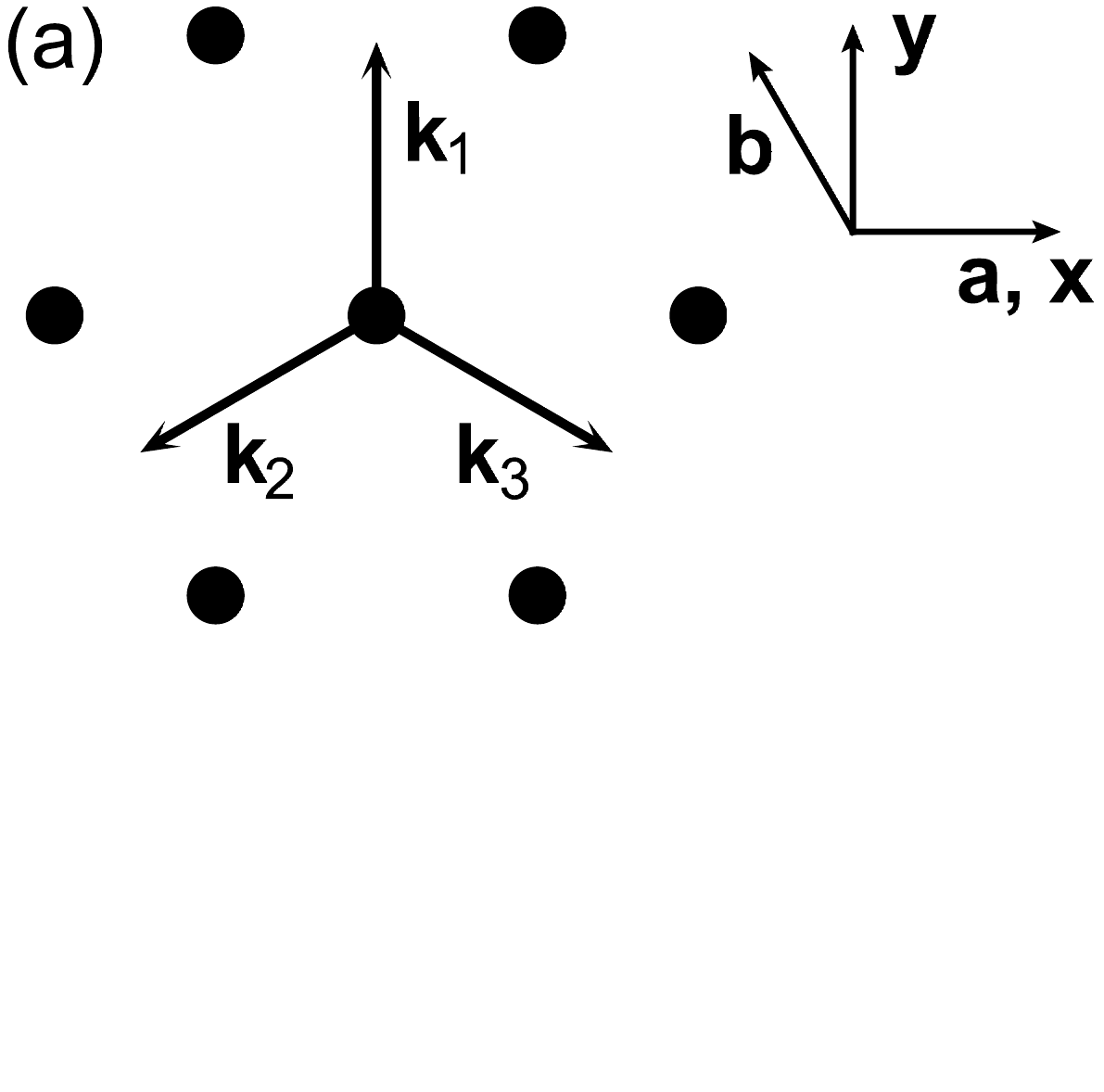}
  \hfill
  \includegraphics[width=4.5cm]{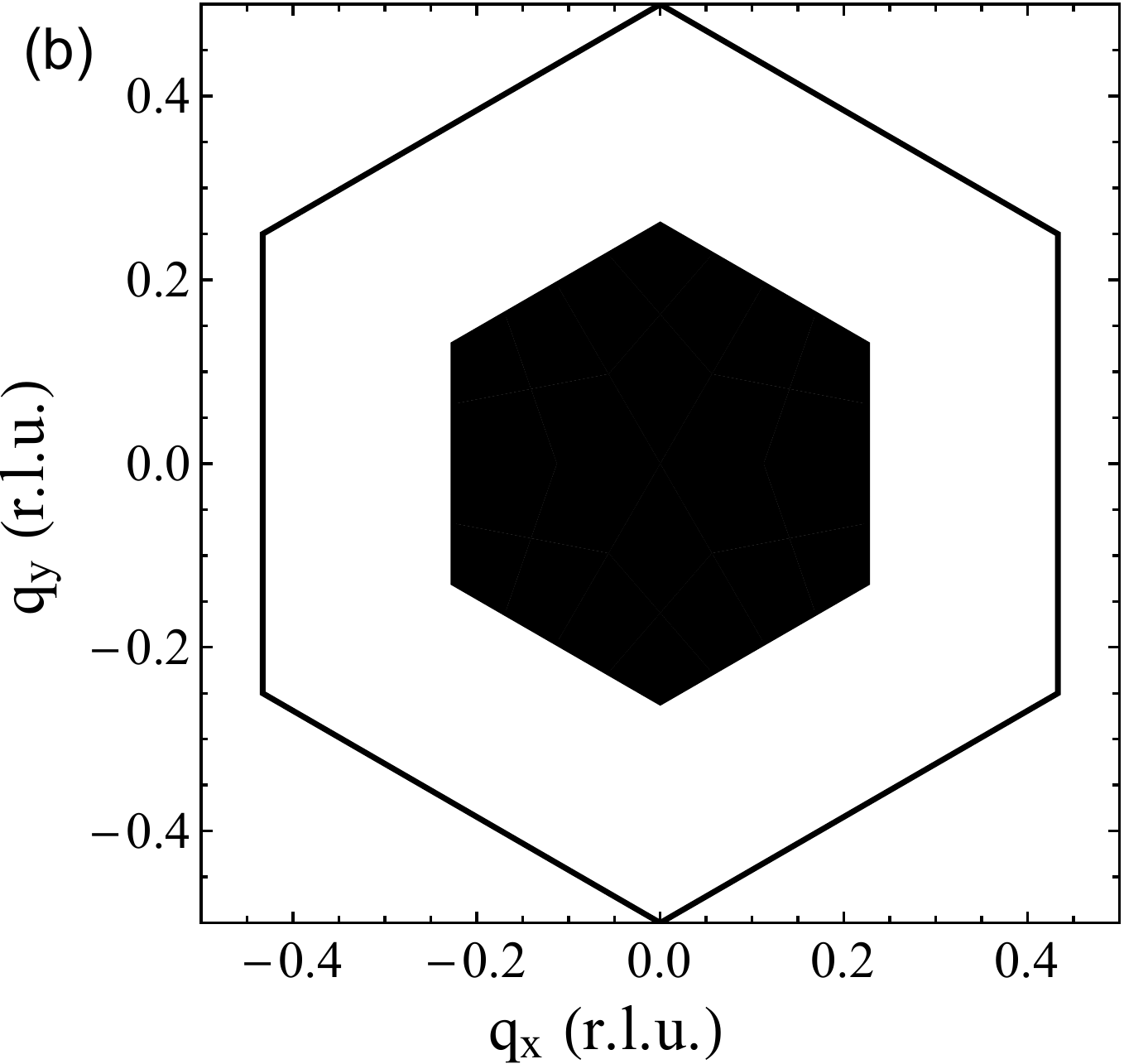}
  \hfill
  \caption{(a) In the considered model, magnetic ions are arranged hexagonally in the \emph{ab} plane. For each incommensurate in-plane modulation vector there are at least two counterparts due to the symmetry of the system. Here we sketch three modulation vectors, which were experimentally observed in Gd$_2$PdSi$_3$~\cite{kurumaji2019SkL}, and show the cartesian $xyz$-coordinates we use in our calculations (out-of-plane $\hat{z}$ direction is chosen along the $\m{c}$ axis). (b) Due to dipolar forces, in the first Brillouin zone (large hexagon) for the in-plane modulation vectors $\m{q}=(q_x,q_y,0)$ the $\m{c}$ axis is the easy one outside the black hexagon, wherein it plays a role of the middle axis; the hard axis is always along the $\m{q}$. Consequently, at relatively small external magnetic fields, the screw helicoids are energetically favorable. Furthermore, we show that in a large part of the phase diagram the triple $Q$ structure is stable.}\label{Fig1}
\end{figure}

Theoretically, stable skyrmions and SkL in frustrated centrosymmetric helimagnets are usually ascribed to interplay of lattice symmetry, exchange interaction, Zeeman energy and certain kind(s) of anisotropic interactions~\cite{okubo2012,leonov2015,lin2016}. For instance, in Ref.~\cite{leonov2015} multiple \emph{Q} states including SkL were predicted at low temperatures for triangular lattice with the easy-axis single-ion anisotropy. Next, it was shown that related models with bilinear and biquadratic $xxz$-type exchange interactions~\cite{hayami2020multiple} and single-ion and bond-dependent anisotropies~\cite{hayami2021noncop} also yields rich phase diagrams in parameter space at low temperature. Furthermore, recent experimental observation of the SkL in tetragonal system  GdRu$_2$Si$_2$~\cite{khanh2020} stimulates theoretical research on low-temperature phases, where importance of biquadratic exchange and compass anisotropy terms were highlighted~\cite{hayami2021square, wang2021}. However, it was shown that the main features of the observed in Ref.~\cite{khanh2020} phase diagram can be described within the simple model with magneto-dipolar interaction and easy-axis anisotropy~\cite{utesov2021tetragonal}.

In the present paper, we show the importance of dipolar forces in the phase diagram of hexagonal frustrated helimagnets, including SkL stabilization. Magneto-dipolar interaction, despite being small, can play significant role in helimagnets properties including temperature- and magnetic-field-induced phase transitions (see, e.g., Refs.~\cite{shiba,gekht1984,gekht,mnbr2,sato,Utesov2017,UtesovMn,utesov2021tetragonal,utesov2021phase}). Noteworthy, for magnetic ions in spherically-symmetrical $L=0$ state (as, e.g., Gd$^{3+}$ in Gd$_2$PdSi$_3$~\cite{kotsanidis1990}) the dipolar forces are of particular importance, since the strength of other anisotropic interactions is governed by the spin-orbit coupling~\cite{white}.

\section{Model}
\label{Smodel}

We consider a simple model of a hexagonal frustrated antiferromagnet with one magnetic ion per crystallographic unit cell. The system Hamiltonian is following:
\begin{eqnarray}
 \label{ham1}
  \mathcal{H} &=& \mathcal{H}_{ex} + \mathcal{H}_{d} + \mathcal{H}_{z}, \nn \\
  \mathcal{H}_{ex} &=& -\frac12 \sum_{i,j} J_{ij} \left(\mathbf{S}_i \cdot \mathbf{S}_j\right), \nn \\
  \mathcal{H}_d &=& \frac12 \sum_{i,j} D^{\alpha \beta}_{ij} S^\alpha_i S^\beta_j,  \\
  \mathcal{H}_z &=& - \sum_i \left(\mathbf{h} \cdot \mathbf{S}_i\right).\nn
\end{eqnarray}
Here, along with the conventional symmetrical Heisenberg and Zeeman interactions [$\m{h}= - g\mu_B \m{H}$ is the external magnetic field in energy units (1~T $\approx$ 1.34~K), which for definiteness will be oriented along the $\m{c}$ axis], we also take into account the dipolar forces. The dipolar tensor reads
\begin{equation}\label{dip1}
	 {\cal D}^{\alpha \beta}_{ij} = \omega_0 \frac{v_0}{4 \pi} \left( \frac{1}{R_{ij}^3} - \frac{3 R_{ij}^\alpha R_{ij}^\beta }{R_{ij}^5}\right),
\end{equation}
where $\alpha$ and $\beta$ denote cartesian coordinates. The strength of the magnetodipolar interaction is governed by
\begin{equation}\label{dipen}
  \omega_0 = 4 \pi \frac{(g \mu_B)^2}{v_0},
\end{equation}
which is usually about $0.1 \div 1$~K ($v_0$ stands for the unit cell volume). For instance, in Gd$_2$PdSi$_3$, one has $\omega_0 \approx 0.53$~K (we use low-temperature lattice parameters from Ref.~\cite{tang2011}).

For the subsequent analysis it is convenient to introduce the Fourier transform
\begin{equation}
\label{four1}
  \mathbf{S}_j = \frac{1}{\sqrt{N}} \sum_\mathbf{q} \mathbf{S}_\mathbf{q} e^{i \mathbf{q} \mathbf{R}_j}.
\end{equation}
Here $N$ is the total number of spins. Plugging this expression into the Hamiltonian~\eqref{ham1} we get
\begin{eqnarray}
  \label{ex2}
  \mathcal{H}_{ex} &=& -\frac12 \sum_\mathbf{q} J_\mathbf{q} \left(\mathbf{S}_\mathbf{q} \cdot \mathbf{S}_{-\mathbf{q}}\right), \\
	\label{dip2}
  \mathcal{H}_d &=& \frac12 \sum_\mathbf{q} {\cal D}^{\alpha \beta}_\mathbf{q} S^\alpha_\mathbf{q} S^\beta_{-\mathbf{q}}. \\
	\label{z21}
 \mathcal{H}_z &=& - \sqrt{N} \left(\mathbf{h} \cdot \mathbf{S}_{\bf 0}\right).
\end{eqnarray}
The former two terms can be combined into bilinear in spin components part
\begin{equation}\label{tens1}
  \mathcal{H}_0 = - \sum_\mathbf{q} \mathcal{H}^{\alpha\beta}_\mathbf{q} S^\alpha_\mathbf{q} S^\beta_{-\mathbf{q}}.
\end{equation}
In the reciprocal space, symmetrical tensor $\mathcal{H}^{\alpha\beta}_\mathbf{q}$ determines three eigenvalues $\lambda_1(\mathbf{q}) \geq \lambda_2(\mathbf{q}) \geq \lambda_3(\mathbf{q})$ corresponding to three mutually perpendicular eigenvectors $\mathbf{v}_1(\mathbf{q}), \, \mathbf{v}_2(\mathbf{q}), \, \mathbf{v}_3(\mathbf{q})$. The latter can be considered as a set of principal axes for momentum-dependent biaxial anisotropy originating from the dipolar interaction. Note, that standard easy-axis and easy-plane anisotropies, and also more tricky compass anisotropy~\cite{banerjee2013,chen2016exotic} can be easily included into this scheme: they simply modify eigenvalues $\lambda_k$.

Since our goal is to describe the high-temperature part of the phase diagram, we denote averaged over thermal fluctuations values $ \langle \m{S}_i \rangle$ as $\m{s}_i$ (and $\langle \m{S}_\m{q} \rangle$ as $\m{s}_\m{q}$). Near the ordering temperature $|\m{s}_i| \ll S$, and one can expand the free energy of the mean-field approach in powers of the order parameters as follows (see, e.g., Refs.~\cite{gekht1984, Utesov2017} for details):
\begin{equation}\label{Free1}
  \mathcal{F} = - \sum_\mathbf{q} \mathcal{H}^{\alpha\beta}_\mathbf{q} s^\alpha_\mathbf{q} s^\beta_{-\mathbf{q}} - \sqrt{N} \mathbf{h} \cdot \mathbf{s}_{\bf 0} + A T \sum_i s^2_i + B T \sum_i s^4_i.
\end{equation}
The expansion constants read~\cite{gekht1984}
\begin{eqnarray}
  A = \frac{3}{2S(S+1)}, \quad
  B = \frac{9[(2S+1)^4-1]}{20 (2S)^4(S+1)^4}.
\end{eqnarray}
For $S=7/2$ one has $A \approx 0.095$ and $B \approx 0.002$.

In general, the ordering temperature of the model~\eqref{Free1} corresponds to the largest eigenvalue $\lambda_1(\m{q})$ where the system becomes unstable towards formation of the magnetic structure with the corresponding momentum. As the dipolar interaction is typically much smaller than the frustrated exchange coupling, this maximum approximately corresponds to the momentum $\m{k}$, which maximizes $J(\m{q})$. It is usually incommensurate due to the frustration, and in low-symmetry lattices either spiral (if $\lambda_1(\m{k})=\lambda_2(\m{k})$) or sinusoidal spin-density wave (SDW) (for $\lambda_1(\m{k}) > \lambda_2(\m{k})$) ordering emerge at $T_c = \lambda_1(\m{k})/A$.

For high-symmetry lattices there can be several equivalent $\m{k}$, which can lead to the stabilization of various so-called multiple-$Q$ structures, and SkL in particular (see, e.g., Ref.~\cite{utesov2021tetragonal} for the discussion of tetragonal frustrated antiferromagnet). Below, we concentrate on a relevant to Gd$_2$PdSi$_3$ case of three equivalent in-plane modulation vectors $\m{k}_1=k(0,1,0),\m{k}_2 = k(-\sqrt{3}/2,-1/2,0), \m{k}_3 = k(\sqrt{3}/2,-1/2,0)$ with angles $120^\circ$ between them (cartesian coordiantes are used, see Fig.~\ref{Fig1}(a) and Ref.~\cite{kurumaji2019SkL}); generalization of the results to another set of in-plane vectors is straightforward.

Importantly, the dipolar tensor for in-plane modulation vectors has quite simple properties. Basically, it favors screw helicoid structure~\footnote{Qualitatively it can be understood using analogy with Bloch domain walls in ferromagnets: spin component along the modulation vector leads to positive correction to the magnetic structure energy from dipolar interaction.}. In more details, the $\m{c}$-axis is a middle axis for relatively small $\m{q}$ and it is an easy axis in the rest part of the first Brillouin zone (see Fig.~\ref{Fig1}(b)). Moreover, the hard axis is approximately parallel to $\m{q}$ with good accuracy (for high-symmetry directions it is an exact feature), so the perpendicular to $\m{q}$ direction plays a role of the easy or middle axis, depending on the $\m{q}$ position in the Brillouin zone [for $\m{k}_{1,2,3}$ we choose the corresponding vectors as follows: $\m{e}_1 = (-1,0,0), \m{e}_2=(1/2,-\sqrt{3}/2,0), \m{e}_3=(1/2,\sqrt{3}/2,0) $]. This axes hierarchy is crucial for the hexagonal SkL stabilization, as it is shown below.

\section{Phase Diagram: in-plane easy axes}

First, we assume that the $\m{c}$ axis is the middle one for the modulation vectors $\m{k}_j$. So, the easy axes are lying in-plane and the angles between them are 120$^\circ$. External magnetic field is applied along the $\m{c}$ axis; corresponding eigenvalue $\lambda_\m{0} = (J_\mathbf{0} - \omega_0 \mathcal{N}_{zz})/2$ ($\mathcal{N}_{zz}$ is the demagnetization tensor component~\cite{SpinWaves} if the shape of the sample is ellipsoid). In order to simplify equations we introduce ``temperature'' $t= \lambda_1 - AT$ (it is positive in the magnetically ordered phases) and parameters $\Lambda = \lambda_1 - \lambda_2$, $\Lambda^{\prime} = \lambda_1 - \lambda_3 > \Lambda$, and $\Lambda_0 = \lambda_1 - \lambda_0$. Finally, it is sufficient to substitute $B T$ by $b = B T_c$ in Eq.~\eqref{Free1}.

Similarly to Ref.~\cite{utesov2021tetragonal}, we restrict our analysis to particular set of magnetic structures. We also neglect possible small variations of the modulation vectors among the phases with multi-component order parameter, which can appear due to dipolar tensor eigenvalues nontrivial momentum-dependence. Details of calculations are mostly presented in the Appendix~\ref{AppendA}.

(i) simple SDW (it will be referred to as 1$S$). One can choose any vector from $\m{k}_{j}$ and arbitrary phase of cosine, e.g.,
\begin{equation}\label{S1S1}
  \mathbf{s}_i = s \mathbf{e}_1 \cos{\mathbf{k}_1 \mathbf{R}_i} + m \hat{z}.
\end{equation}
Corresponding free energy per one spin reads
\be \label{F1S}
  \mathcal{F}_{1S} &=& -\frac{t}{2}s^2 - h m - (t-\Lambda_0) m^2 + \nn \\
   &&+ b \left(m^4 + m^2 s^2 +\frac{3}{8} s^4 \right).
\ee
(ii) simple helicoid with spins rotating in the easy plane perpendicular to $\m{q}$ (it will be referred to as 1$Q$). Taking, e.g., $\m{k}_1$ we have
\begin{equation}\label{S1Q1}
  \mathbf{s}_i = s \mathbf{e}_1 \cos{\mathbf{k}_1 \mathbf{R}_i} + p \hat{z} \sin{\mathbf{k}_1 \mathbf{R}_i} + m \hat{z}.
\end{equation}
Here $s p>0$ corresponds to the right spiral, and $s p <0$ to the left one, also the common phase of sine and cosine functions can be chosen arbitrary. Corresponding free energy per one spin reads
\be \label{F1Q}
  \mathcal{F}_{1Q}= -\frac{t}{2}s^2 - \frac{t-\Lambda}{2}p^2- h m - (t-\Lambda_0) m^2 + \nn \\
    b \left[m^4 + m^2 (s^2 + 3 p^2) +\frac{3s^4 + 2 s^2 p^2 + 3 p^4}{8}  \right].
\ee
(iii) conical spiral with spins rotating in the \emph{ab} plane, perpendicular to the magnetic field (this structure will be referred to as XY). For spin ordering one has
\begin{equation}\label{SXY1}
  \mathbf{s}_i = s \mathbf{e}_1 \cos{\mathbf{k}_1 \mathbf{R}_i} + p \hat{z}\times\mathbf{e}_1 \sin{\mathbf{k}_1 \mathbf{R}_i} + m \hat{z}.
\end{equation}
Once again, signs of $s$ and $p$ can be taken arbitrary as well as the common phase.
Corresponding free energy per one spin reads
\be \label{FXY}
  \mathcal{F}_{XY}= -\frac{t}{2}s^2 - \frac{t-\Lambda^\prime}{2}p^2- h m - (t-\Lambda_0) m^2 + \nn \\
    b \left[m^4 + m^2 (s^2 + p^2) +\frac{3s^4 + 2 s^2 p^2 + 3 p^4}{8}  \right].
\ee
Important differences with Eq.~\eqref{F1Q} are the following: $\Lambda^\prime$ instead of $\Lambda$ and $b m^2 p^2$ instead of $3 b m^2 p^2$, so at low magnetic fields the 1$Q$ structure is preferable, but at stronger fields $\mathcal{F}_{XY}$ becomes smaller.

(iv) the triple-$Q$ structure, which is a superposition of three screw helicoids (3$Q$). In general, one can consider this ordering with many parameters
\be\label{S3Q1}
  \mathbf{s}_i &=& \sum_{j=1,2,3} \left[s_j \mathbf{e}_j \cos{(\mathbf{k}_j \mathbf{R}_i + \varphi_j)} + p_j \hat{z} \sin{(\mathbf{k}_j \mathbf{R}_i + \varphi_j)}\right] \nn \\ &&+ m \hat{z}.
\ee
However, a minimal free energy can be achieved only if [except for the 1$Q$ structure, which can be also described by Eq.~\eqref{S3Q1}] $s_1=s_2=s_3=s/\sqrt{3}$, $p_1=p_2=p_3 = p/\sqrt{3}$, and the following restriction on phases is satisfied: $\varphi_1+\varphi_2+\varphi_3 = 2 \pi n + \textrm{sign}(p) \pi/2$.~\footnote{Note, that the chiralities of all three helicoids are the same.} Corresponding free energy is given by
\be \label{F3Q}
  \mathcal{F}_{3Q}= -\frac{t}{2}s^2 - \frac{t-\Lambda}{2}p^2- h m - (t-\Lambda_0) m^2 + \nn \\
    b \Biggl[m^4 + m^2 (s^2 + 3 p^2) +\frac{9s^4 + 10 s^2 p^2 + 15 p^4}{24} \\ - \frac{m p (2p^2+s^2)}{\sqrt{3}} \Biggr]. \nn
\ee
The last term here is the most important one. Note, that if all $p_j=0$ in Eq.~\eqref{S3Q1} the corresponding magnetic structure is a triple SDW or 3$S$. Moreover, its free energy is the same with the simple 1$S$ structure~\footnote{This degeneracy can be lifted by taking into account $\propto s^6$ terms in the free energy. The result is that the 1$S$ free energy is always slightly smaller than the 3$S$ one}. However, it is evident from Eq.~\eqref{F3Q} that even at infinitesimal $h$, when nonzero $m$ appears, the 3$S$ structure is unstable towards transition to 3$Q$ structure. So, if the easy axes are in-plane, the 3$Q$ structure always has lower energy than the 1$S$ and 3$S$.

For analytical treatment of these phases free energies and phase diagram description, we use the following trick: in equation $\partial \mathcal{F}/\partial m = 0$ we neglect terms with the forth power of order parameter components, assuming that the system near $T_c$ is not close to ferromagnetic instability. So, the magnetization simply reads
\be \label{mag1}
  m(t,h) = \chi(T) h = \frac{h}{2(\Lambda_0-t)}.
\ee
Furthermore, usually $\Lambda_0 \gg \Lambda, \Lambda^\prime$ and it is sufficient to put $\chi \equiv \chi(T_c) = 1/ 2 \Lambda_0$ instead of $\chi(T)$. Under this assumption, it is fruitful to use parameters $t$ and $\Lambda$ renormalized by terms $\propto m^2$ in calculations, which can be defined as
\be \label{th}
  t_h = t - 2 b (\chi h)^2, \\ \label{Lh}
  \Lambda_h = \Lambda + 4 b (\chi h)^2.
\ee
Then, it is easy to show that both 1$S$ and 3$Q$ phases require $t_h>0 \Leftrightarrow t> 2 b (\chi h)^2$. This inequality determines the phase boundary between 3$Q$ and paramagnetic (or induced ferromagnetic) phase, PM.

If we for a moment forget about the 3$Q$ structure a simple phase diagram shown in Fig.~\ref{FPhaseWO3Q} can be easily obtained. Important scales here are determined by the coordinates of the triple point (where 1$S$, 1$Q$, and XY are in equilibrium), namely the spiral plane flop field
\be \label{hsf}
  h_{SF} = \sqrt{\frac{\Lambda^\prime - \Lambda}{4 b \chi^2}}
\ee
and the ``temperature''
\be \label{ttr}
  t_{Tr} = 2 \Lambda^\prime - \Lambda/2.
\ee

\begin{figure}
  \centering
  \includegraphics[width=6cm]{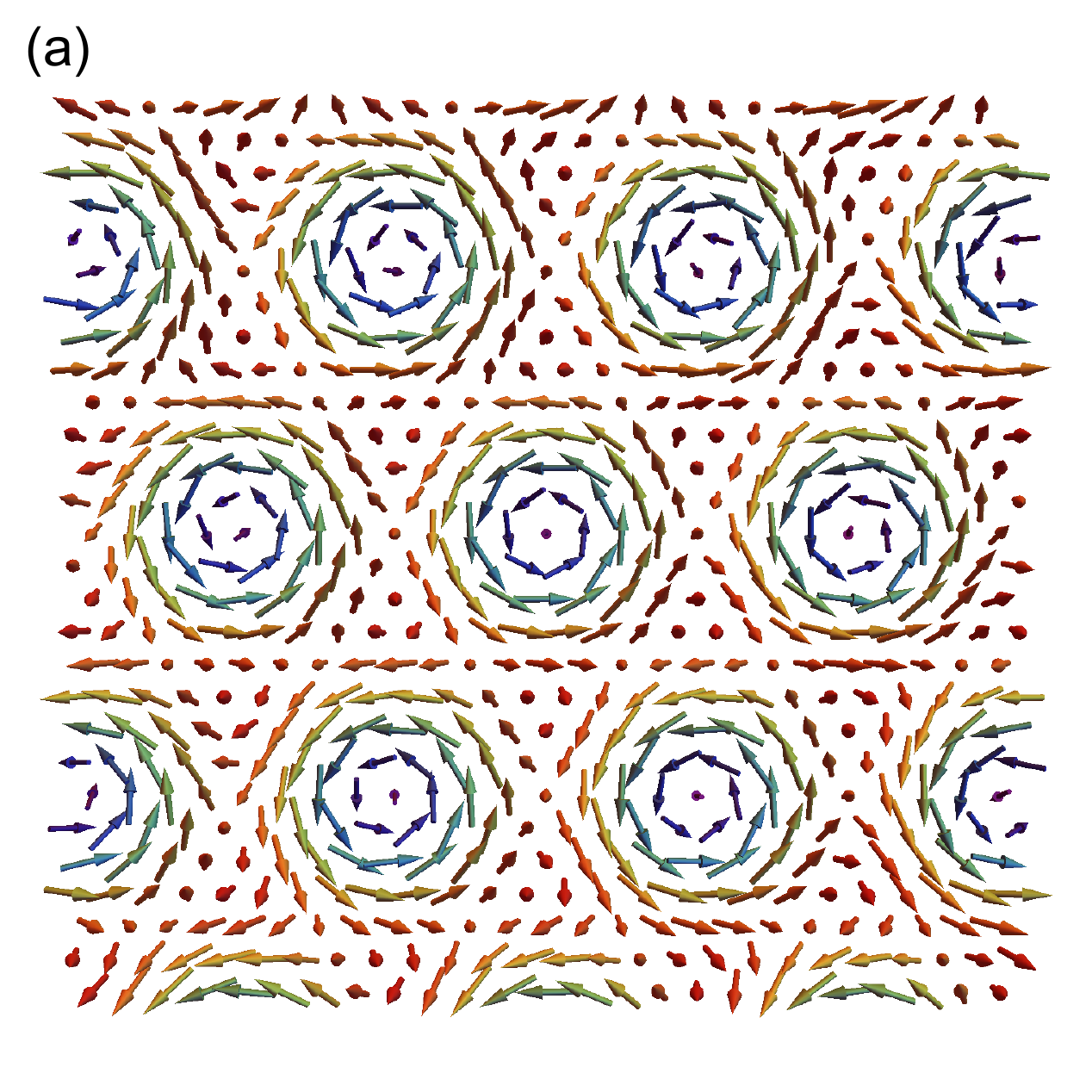}
  \centering
  \includegraphics[width=6cm]{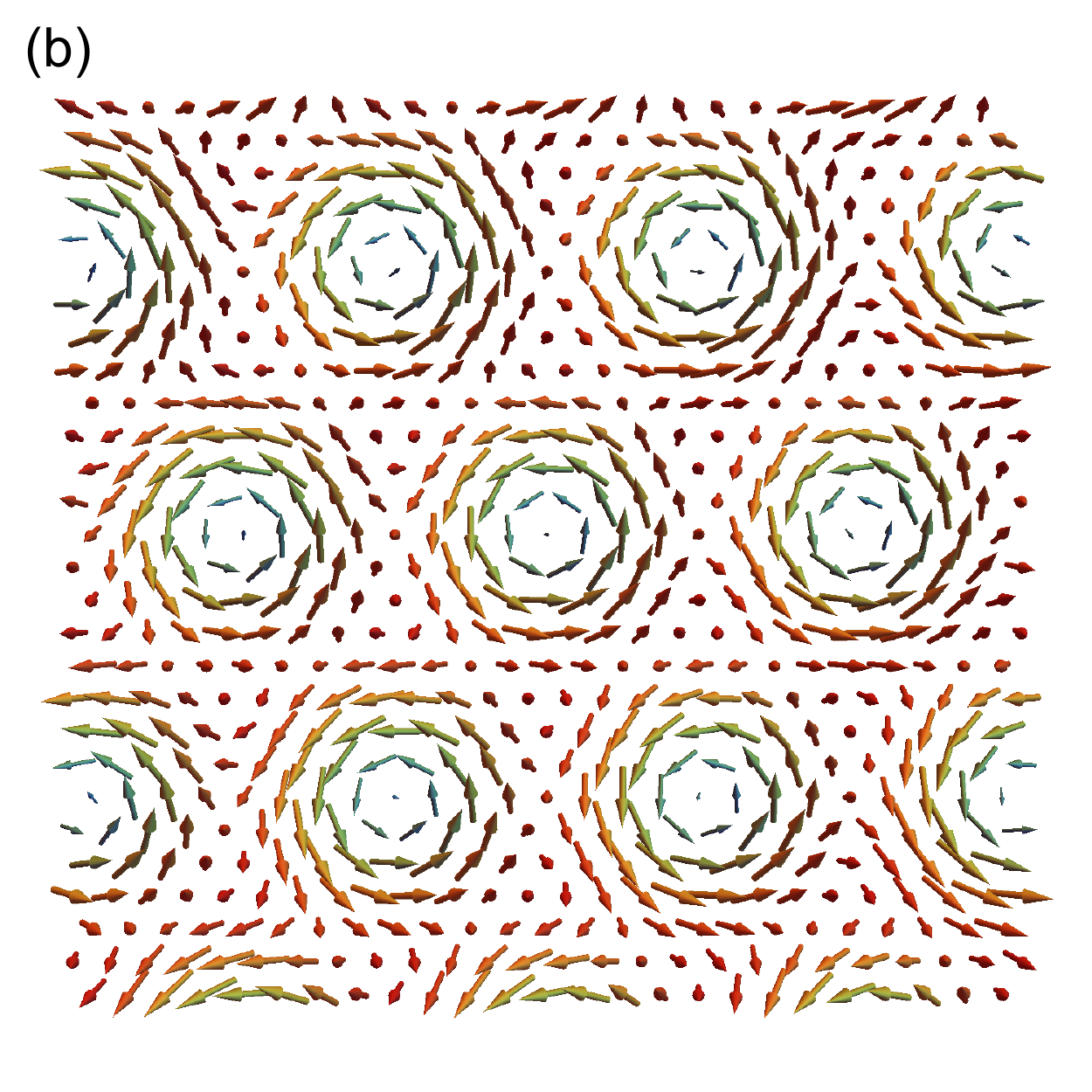}
  \caption{The 3$Q$ phase, which is a superposition of three homochiral (left or right) helicoids and constant spin component -- magnetization, can be either topologically nontrivial at moderate magnetic fields (a), where spins wrap around a sphere ones per skyrmion, or trivial (b) at fields close to the saturation one (see Fig.~\ref{FPhase1}). In the latter case, $z$ components of all spins are positive (however, with non-negligible modulated part) and the corresponding topological charge is evidently zero. Here $\m{h} \uparrow\uparrow \m{c}$, red spins are ``up'' ($s_z>0$) and violet spins are ``down'' ($s_z<0$) . }\label{FigSkL}
\end{figure}

However, 3$Q$ has lower energy than 1$S$, and it pushes the XY phase further away to $t$ substantially larger, than $t_{Tr}$. Moreover, below we show that for realistic parameters of Gd$_2$PdSi$_3$ the XY phase does not appear at all in the range of mean-field approach validity. So, the resulting phase diagram consists of 1$Q$ and 3$Q$ phases. The boundary between them is approximately given by
\be \label{t1Q3Q}
  t_{1Q-3Q}(h) \approx \frac{3}{2} \Lambda + 45 b (\chi h)^2.
\ee
Importantly, inside 3$Q$ phase there exists a boundary dividing topologically trivial and nontrivial parts. We show these two spin structures in Fig.~\ref{FigSkL}. For nontrivial structure topology with $n_{Sk} = \pm 1$ (the sign here depends on chirality of helicoid constituents), spin should be able to point against the magnetization, which is equivalent to $ \sqrt{3} p > m$. In our approximation this boundary can be found exactly as
\be \label{tSkL}
  t_{SkL}(h) = \frac{9}{10} \Lambda + \frac{203}{45} b (\chi h)^2 \approx 0.9 \Lambda + 4.5 b (\chi h)^2.
\ee
At $t <t_{SkL}$ (which corresponds to larger temperatures in reality) the 3$Q$ structure is topologically trivial with $n_{Sk} = 0$. Note, that in our approach $t_{SkL}(h)$ has not physical meaning of a phase transition temperature. However, one can expect some anomalies due to possible effects of topology on other, e.g., electronic degrees of freedom.

\begin{figure}
  \centering
  \includegraphics[width=8cm]{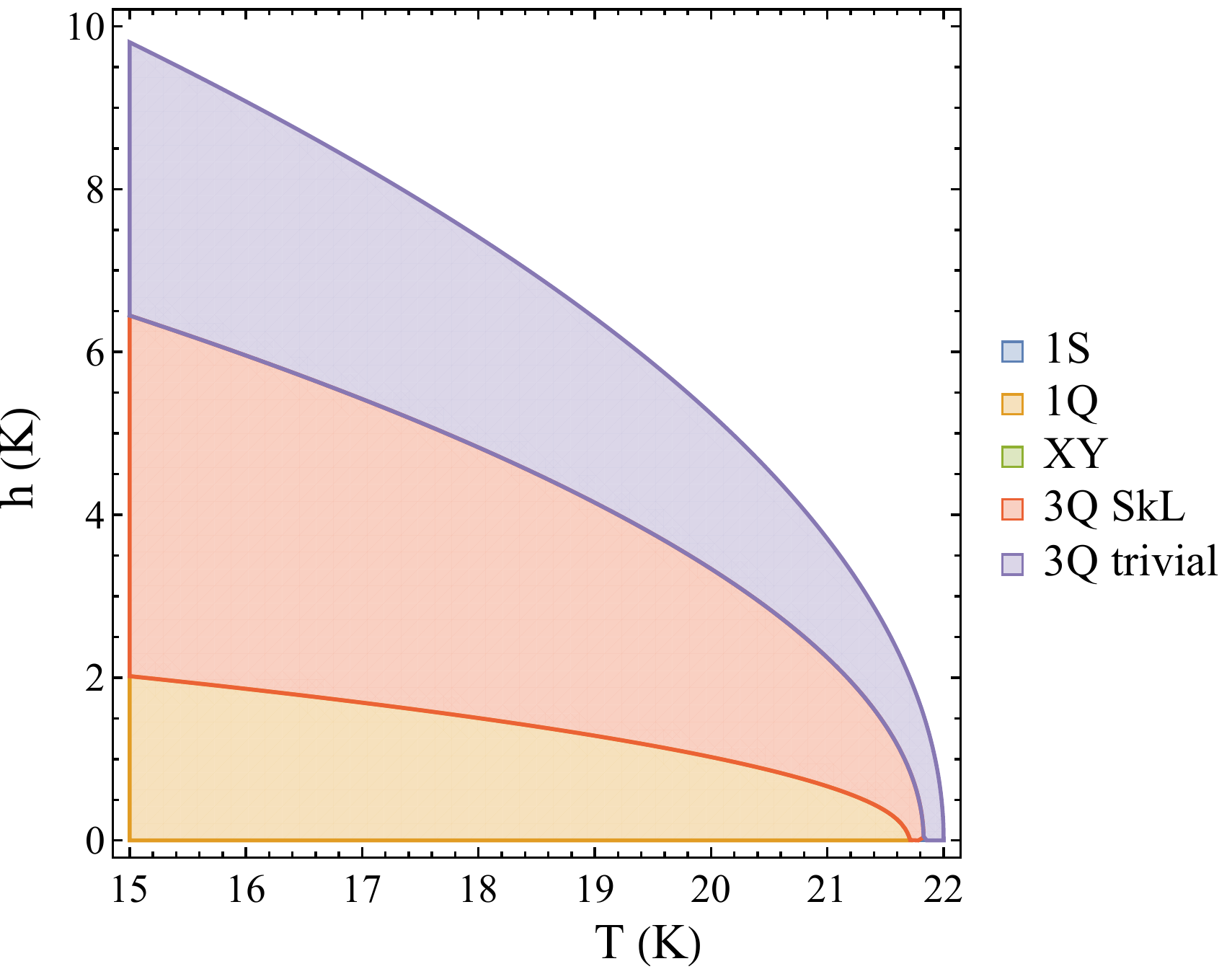}\\
  \caption{High-temperature part of the phase diagram for centrosymmetric hexagonal frustrated antiferromagnet with dipolar interaction (see Fig.~\ref{Fig1}). The parameters~\eqref{param1} relevant to Gd$_2$PdSi$_3$ were used. Usual for dipolar forces-induced anisotropy simple SDW (1S) and conical helicoid (XY) (see Fig.~\ref{FPhaseWO3Q}) do not appear in this region of the phase diagram. They are substituted by the 3$Q$ phase in its topologically trivial and non-trivial forms (see text). }\label{FPhase1}
\end{figure}

We apply our theory to Gd$_2$PdSi$_3$, which parameters can be estimated using the ordering temperature $T_N \approx 22$~K and the saturation field $H_S \approx 9$~T~\cite{hirschberger2020, spachman2021}. The former quantity determines $J_{\m{k}}$ for $k=0.14$~\cite{kurumaji2019SkL} (in reciprocal lattice units, r.l.u.), whereas the second one can be used to estimate $J_{\m{0}}$, since $h_S \approx S(J_{\m{k}} - J_{\m{0}})$ in frustrated helimagnets (see, e.g., Refs.~\cite{Nagamiya1962,utesov2021phase}). Dipolar tensor components can be calculated using their representation in the form of fast converging sums (see, e.g., Ref.~\cite{cohen}). As a result we get (all values are in kelvins)
\be \label{param1}
  J_{\m{k}} \approx 4.00, \, J_{\m{0}} \approx 0.56, \nn \\
  \Lambda \approx 0.02, \, \Lambda^\prime \approx 0.26, \, \Lambda_0 \approx 1.72.
\ee
Using this set of parameters we obtain the phase diagram shown in Fig.~\ref{FPhase1}. Note, that it is similar to the high-temperature part of the phase diagram, observed experimentally in Refs.~\cite{kurumaji2019SkL,hirschberger2020} for Gd$_2$PdSi$_3$ compound. Our theory suggests topologically trivial 3$Q$ structure for IC-2 (however, its more complicated that the 3$S$ vortex structure with constant $s_z$, which was described in Ref.~\cite{gekht1984}), Bloch-type hexagonal SkL for the $A$ phase, and simple 1$Q$ helicoid phase for IC-1 (notation IC-1, IC-2 and $A$ phase belong to Ref.~\cite{kurumaji2019SkL}). We would like to stress, that triangular meron-antimeron lattice which is 3$Q$ structure with $\varphi_1 + \varphi_2 + \varphi_3 = \pi n$ in our high temperature approach always has larger free energy than 1$Q$, see Eq.~\eqref{AF3Q1}. IC-1 phase non-coplanarity~\cite{kurumaji2019SkL} can be thus a result of three helical domains and the domain walls among them. It can be checked experimentally by applying in-plane magnetic field along certain $\m{k}_j$, which will choose the corresponding helical domain. However, we note, that meron-antimeron lattices can be stabilized in systems with biquadratic exchange and Dzyaloshinskii-Moriya interaction~\cite{hayami2021meron}.

The type of the phase diagram shown in Fig.~\ref{FPhase1} is quite general (in a qualitative sense) for the model with in-plane easy axes. Nevertheless, by varying the parameters, XY phase can appear in the approach validity region. We illustrate this statement by manually setting $\Lambda^\prime = 0.05$ in Eq.~\eqref{param1}, which results in the phase diagram shown in Fig.~\ref{FPhaseXY1} with observable regions of the conical phase. Note, that in this case $T_{Tr} \approx 21$~K, however XY emerges only at $T \approx 18$~K due to the competition with 3$Q$.

\section{Easy axis along $\m{c}$}

According to Fig.~\ref{Fig1} there is a possibility to have collinear easy axes along $\m{c}$  for all three $\m{k}_j$ solely due to dipolar interaction. However, this requires rather large $k$. At the same time, standard single-ion easy-axis anisotropy can also change the axes hierarchy (its constant should be subtracted from the $\mathcal{H}^{\alpha \beta}_\m{q}$ in-plane eigenvectors eigenvalues and added to the ones corresponding to the $\m{c}$ axis). In both cases we can arrive to substantially different phase diagram, which is discussed below.

First, we would like to point out several important differences with the previous case: (i) modulated component of the 1$S$ phase is now along $\hat{z}$, (ii) for free energies of 1$Q$ and 3$Q$ phases $t-\Lambda$ is multiplied on $s^2$, not on $p^2$ as previously, (iii) in conical XY phase modulated spin component rotates perpendicular to the easy axis, and (iv) in magnetic field the following counterparts of Eqs.~\eqref{th} and~\eqref{Lh} should be used:
\be \label{th2}
  t_h = t - 6 b (\chi h)^2, \\ \label{Lh2}
  \Lambda_h = \Lambda - 4 b (\chi h)^2.
\ee

\begin{figure}
  \centering
  \includegraphics[width=6cm]{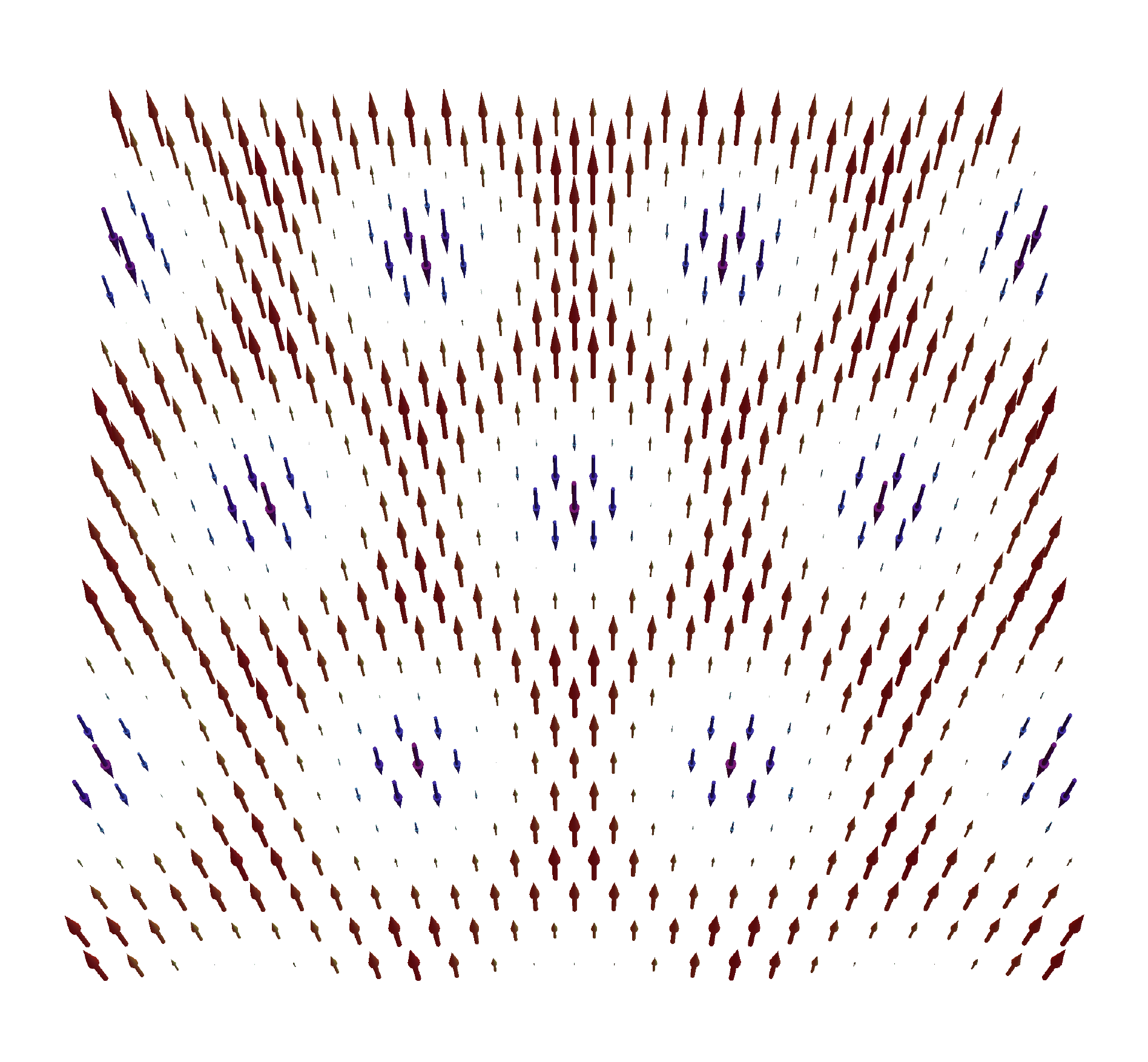}\\
  \caption{When the easy axes are along the $\m{c}$, peculiar 3$P$ phase can be stable in the external field near $T_c$ . It consists of three collinear spin-density waves (see text). }\label{Fig3P}
\end{figure}

Importantly, here the in-plane $s$-component of the 3$Q$ as well as 1$Q$ structure appears only at $t > 3 \Lambda/2$ (see Appendix~\ref{AppendB} for the details). At lower $t$ the 1$S$ phase competes with the phase we dub 3$P$, which is the superposition of three collinear spin-density waves~\footnote{The corresponding spin structure is given by Eq.~\eqref{S3Q1} with all $s_j=0$, $p_1=p_2=p_3=p/\sqrt{3}$ and $\varphi_1+\varphi_2+\varphi_3 = 2 \pi n + \textrm{sign}(p) \pi/2$, see Fig.~\ref{Fig3P}.}. If we fix certain $t< 3 \Lambda/2$ and increase $h$ starting from $h=0$ we have sequence of phase transitions $1S \rightarrow 3P \rightarrow \text{PM}$. Both transitions are of the first order.

At larger ``temperatures'' $t > 3 \Lambda/2$ (lower $T$), this sequence transforms into $1Q \rightarrow 3Q \rightarrow \text{PM}$. The former transition is always the first order one; however the latter can be either continuous or discontinuous. The reason is that in a small vicinity of $t=3 \Lambda/2$ the 3$Q$ phase is topologically-nontrivial in the whole range of fields till PM phase becomes the ground state. For yet larger $t$, before PM phase appears, the 3$Q$ phase becomes topologically trivial (either smoothly or discontinuously, see Appendix~\ref{AppendB}).

\begin{figure}
  \centering
  \includegraphics[width=8cm]{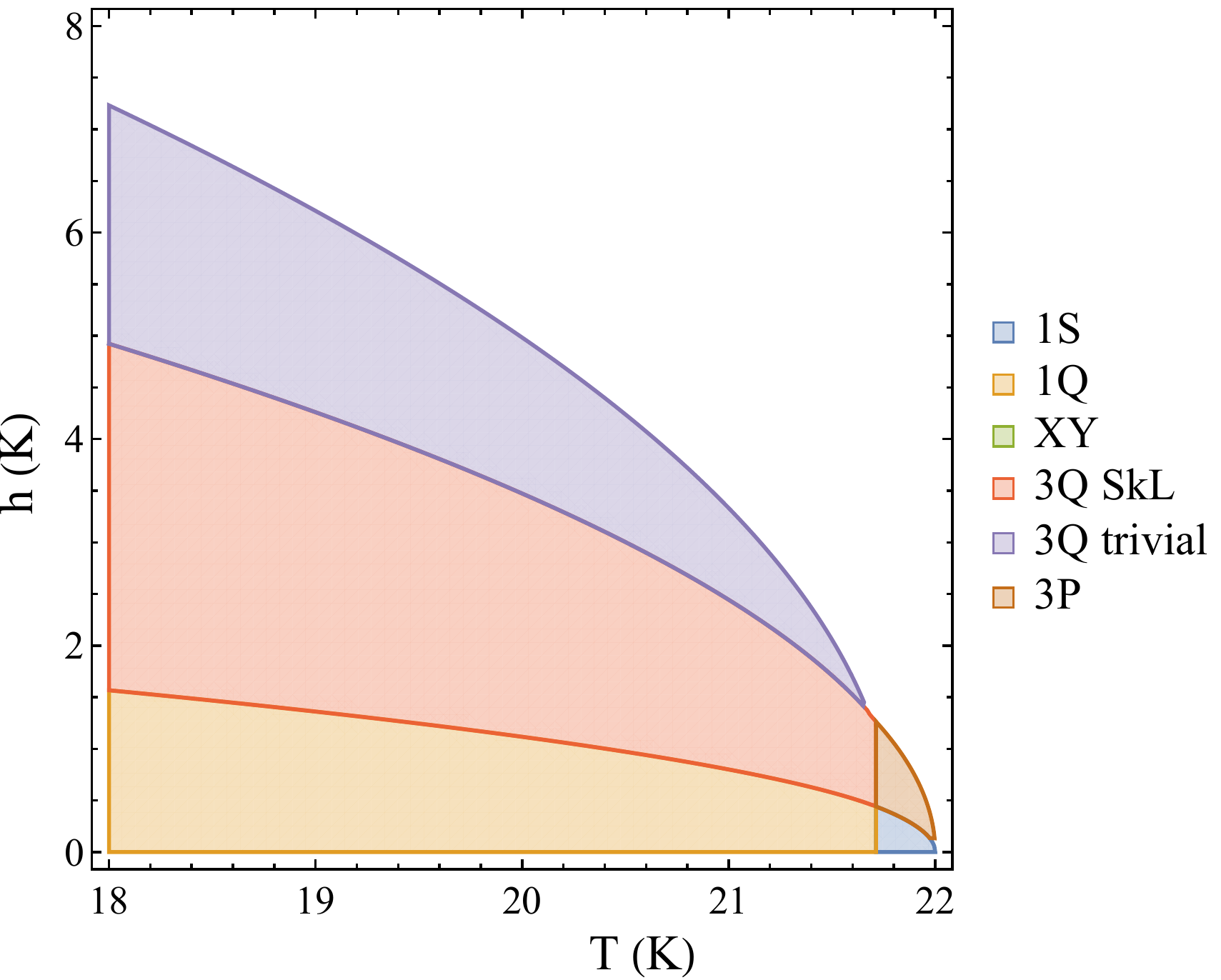}\\
  \caption{The same as in Fig.~\ref{FPhase1}, but for the system with collinear easy axes along $\m{c}$ for all modulation vectors $\m{k}_j$. Parameters~\eqref{param1} are used. Note, that the superposition of three collinear spin-density waves, 3$P$ phase (see Fig.~\ref{Fig3P}), appears.}\label{FPhase2}
\end{figure}

As a result, using parameters set~\eqref{param1} (note, that if we keep the same $\m{k}_j$ as in previous part of the paper it implies single-ion easy-axis anisotropy with the constant equal to $\Lambda$, which makes $\m{c}$ the easy direction) we obtain the phase diagram shown in Fig.~\ref{FPhase2}. The boundary between 1$S$ (1$Q$) and 3$P$ (3$Q$) is approximately given by
\be
  t_{1S-3P} \approx 40 b(\chi h)^2,
\ee
whereas the boundary between 3$P$ and PM reads
\be
  t_{3P-PM} \approx \frac{74}{15} b(\chi h)^2.
\ee
The 3$Q$ Skl -- 3$Q$ trivial boundary is approximately the previous equation continuation onto larger $t$:
\be \label{tSkL2}
  t_{SkL} = \frac{\Lambda}{10} + \frac{203}{45} b (\chi h)^2.
\ee
Finally, the second order phase transition between 3$Q$ trivial and PM takes place at
\be \label{t3QPM}
  t_h=\Lambda_h \leftrightarrow t = \Lambda + 2 b (\chi h)^2.
\ee
Presented above curves fix the generic type of phase diagrams which can be obtained by varying parameters of the system with the condition that $\m{c}$ axis remains the easy one for modulated spin components.

\begin{figure}
  \centering
  \includegraphics[width=8cm]{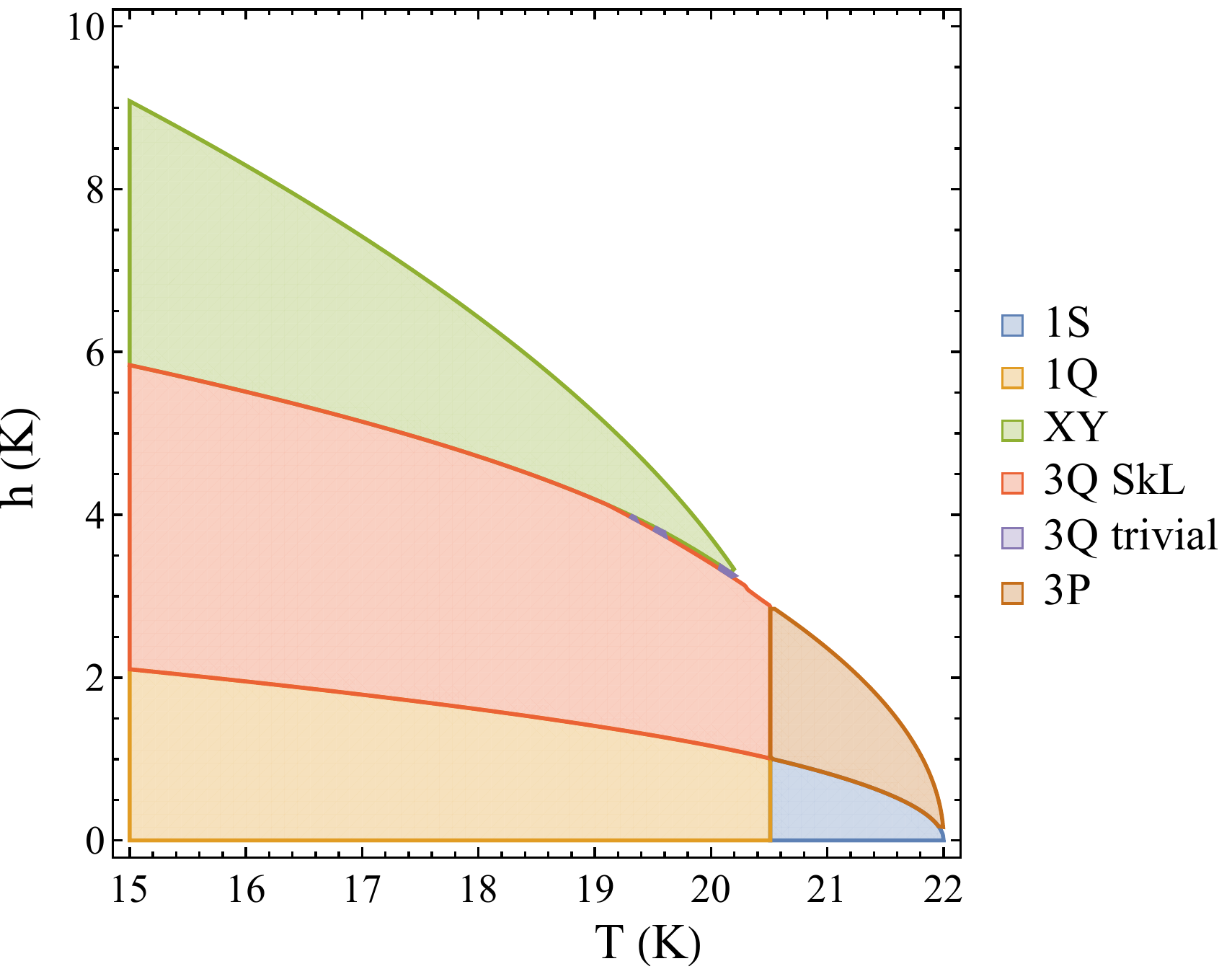}\\
  \caption{If one considers standard single-ion easy-axis anisotropy without dipolar forces, the conical phase becomes stable at large fields (see text), but the skyrmion lattice is the ground state in the intermediate fields range. In contrast, for easy-plane anisotropy the whole phase diagram is rather trivial, only XY and PM can be observed.}\label{FPhase3}
\end{figure}

It is pertinent to make a connection with the paper~\cite{leonov2015} devoted to low temperatures and consider standard easy-plane and easy-axis anisotropies. In agreement with the results of Ref.~\cite{leonov2015} we find that the phase diagram for easy-plane anisotropy is trivial and consists only of XY and PM phases. The easy-axis case is more interesting. To analyze it we put $\Lambda = \Lambda^\prime=0.1$~K in the parameters set~\eqref{param1} and observe the phase diagram shown in Fig.~\ref{FPhase3}, which is similar with the one for dipolar forces and easy-axis anisotropy (see Fig.~\ref{FPhase2}) but with the conical XY phase neighboring to PM instead of 3$Q$ trivial one.

Finally, we would like to point out that the complementary to proposed here theory for low-temperature part of the phase diagram is an interesting and challenging problem due to multi-harmonic skyrmion structure (see, e.g., Ref.~\cite{timofeev2021}) and long-range nature of the dipolar forces.

\section{Conclusions}
\label{Sconc}

To conclude, we propose a simple analytical mean-field approach for skyrmion lattices in hexagonal frustrated antiferromagnets capable to describe high-temperature part of the phase diagram. We show, that dipolar forces (which are always present in real compounds) are sufficient to stabilize the SkL for the case of in-plane modulation vectors. We observe several generic types of phase diagrams and discuss the phase boundaries. One of the obtained phase diagrams can be relevant to the experimental observations in Gd$_2$PdSi$_3$~\cite{kurumaji2019SkL,hirschberger2020}.

\begin{acknowledgments}

We are grateful to V.A.\ Ukleev for valuable discussions. The reported study was funded by the Russian Federation President Grant No. MK-1366.2021.1.2.

\end{acknowledgments}

\appendix

\section{Free energies of various phases and phase diagram for in-plane easy axes}
\label{AppendA}

Here we derive analytical expressions for relevant phases free energies. For magnetization components of the corresponding spin structures we use simple approximation $m = \chi h$  with constant susceptibility parameter $\chi$ (see Eq.~\eqref{mag1} and the discussion below). Finally, we make conclusions about the generic phase diagram for in-plane easy axes.

\subsection{Simple spin-density wave (1$S$ phase)}

In this case there is only one component of the order parameters. It can be easily found from Eq.~\eqref{F1S} by plugging $m= \chi h$ and using $t_h$ defined in Eq.~\eqref{th}, which yields
\be \label{AS1S1}
 s^2 &=& \frac{2t_h}{3b}, \\
 \label{AF1S1} \mathcal{F}_{1S} &=& -\frac{t_h^2}{6b} - \frac{\chi h^2}{2}, \quad t_h>0.
\ee
Condition $t_h=0 \Leftrightarrow t = 2 b (\chi h)^2$ determines the boundary between 1$S$ and high field paramagnetic or induced ferromagnetic (PM) phases.

\subsection{Single-$Q$ elliptical helicoid (1$Q$ phase)}

This structure is characterized by two parameters, which measure the amplitude of the spin ordering along the easy and the middle axes. Minimization of the free energy~\eqref{F1Q} yields
\be \label{AS1Q1}
  s^2 &=& \frac{2 t_h + \Lambda_h}{4b}, \nn \\
  p^2 &=&  \frac{2 t_h - 3 \Lambda_h}{4b}
\ee
and
\be \label{AF1Q1}
  \mathcal{F}_{1Q} = - \frac{4 t^2_h - 4 t_h \Lambda_h + 3 \Lambda^2_h}{16 b} - \frac{\chi h^2}{2}, \, t_h> \frac{3}{2}\Lambda_h .
\ee
At
\be \label{At1Q}
t_h= 3 \Lambda_h/2 \Leftrightarrow t_{1Q} = 3 \Lambda/2 + 8 b (\chi h)^2
\ee
continuous transition between 1$S$ and 1$Q$ takes place.

\subsection{Conical helicoid (XY phase)}

\begin{figure}[b]
  \centering
  \includegraphics[width=8cm]{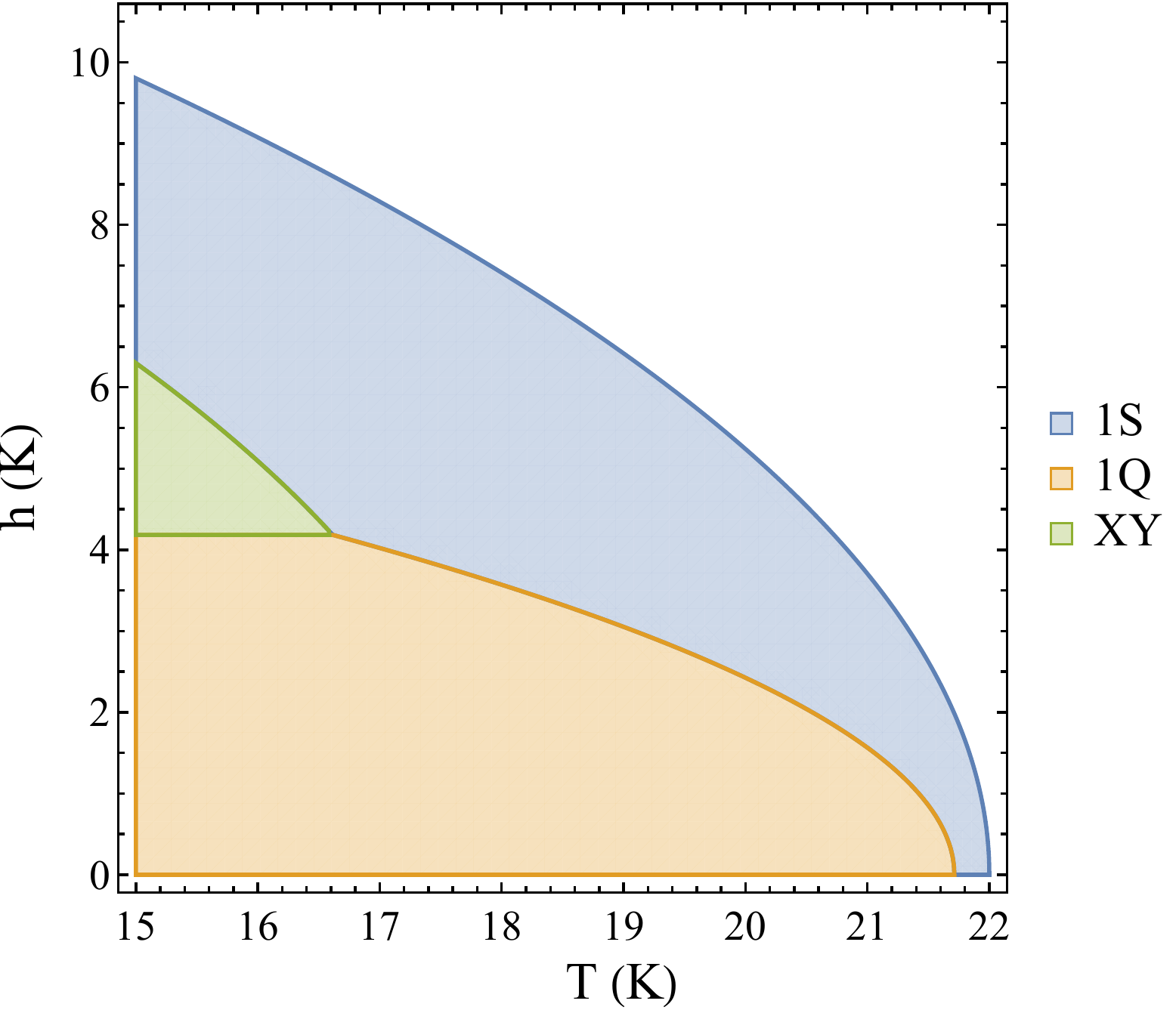}\\
  \caption{High-temperature part of the phase diagram for the considered model with in-plane easy axes, where (for illustration purposes) the 3$Q$ phase was excluded from the analysis. Parameters~\eqref{param1} were used. }\label{FPhaseWO3Q}
\end{figure}

Here modulated spin component rotates in the $ab$ plane, which is perpendicular to the external field. The equations are quite similar with the 1$Q$ phase ones:
\be \label{ASXY1}
  s^2 &=& \frac{2 t_h + \Lambda^\prime}{4b}, \nn \\
  p^2 &=&  \frac{2 t_h - 3 \Lambda^\prime}{4b},
\ee
whereas the free energy reads
\be \label{AFXY1}
  \mathcal{F}_{XY} = - \frac{4 t^2_h - 4 t_h \Lambda^\prime + 3 {\Lambda^\prime}^2}{16 b} - \frac{\chi h^2}{2}, \, t_h> \frac{3}{2}\Lambda^\prime.
\ee
Condition $t_h= \frac{3}{2}\Lambda^\prime \Leftrightarrow t = 3 \Lambda^\prime/2 + 2 b (\chi h)^2$ determines the phase boundary between 1$S$ and XY phases. Moreover, it is evident from Eqs.~\eqref{AF1Q1} and~\eqref{AFXY1} that 1$Q$ and XY phases are in equilibrium when $\Lambda_h = \Lambda^\prime$. This condition determines the so-called spiral plane flop field
\be \label{AhSF1}
 h_{SF} = \sqrt{\frac{\Lambda^\prime - \Lambda}{4 b \chi^2}}.
\ee
Note, that at this field and $t_h = 3 \Lambda_h/2 = 3\Lambda^\prime/2$ phases 1$S$, 1$Q$ and XY are in equilibrium. This yields the triple point temperature
\be
  t_{Tr} = 2 \Lambda^\prime - \Lambda/2.
\ee

If we for the moment forget about the 3$Q$ phase, which properties are described below, the typical phase diagram of the model~\eqref{Free1} considered here is shown in Fig.~\ref{FPhaseWO3Q}, where the parameters set~\eqref{param1} is used. This type of the phase diagram should be contrasted to those (see Figs.~\ref{FPhase1} and~\ref{FPhaseXY1}), where 3$Q$ phase is also taken into account. Note, that for the parameters~\eqref{param1} at $t \lesssim t_{Tr}$ small $|\m{s}_i| \ll S$ expansion is inapplicable: using Eq.~\eqref{AS1S1} one obtains $s \approx 2.5$ for $t = t_{Tr}$ and $h=h_{SF}$.

\subsection{Superposition of three screw helicoids (3$Q$ phase)}

In this case the free energy is a function of three $s_j$, three $p_j$, and three phases $\varphi_j$ (see Eq.~\eqref{S3Q1}). The general expression for it is quite cumbersome, to make it shorter we introduce $s^2_\Sigma=s^2_1+s^2_2+s^2_3, \, p^2_\Sigma = p^2_1+p^2_2+p^2_3$. As a result we get:

\begin{widetext}

\be \label{AF3Q1}
  \mathcal{F}_{3Q} &=& -\frac{t}{2}s^2_\Sigma - \frac{t-\Lambda}{2}p^2_\Sigma - h m - (t-\Lambda_0) m^2 + b \Biggl[ m^4 + m^2 \left( s^2_\Sigma +
  3 p^2_\Sigma \right) + \frac{3(s^2_\Sigma)^2 +2s^2_\Sigma p^2_\Sigma + 3(p^2_\Sigma)^2}{8} \nn \\
  && + \frac{3(p^2_1p^2_2+p^2_1p^2_3+p^2_2p^2_3)}{4} + \frac{s^2_1(p^2_2+p^2_3)+s^2_2(p^2_1+p^2_3)+s^2_3(p^2_1+p^2_2)}{4} \\
  && - m(6 p_1 p_2 p_3 + s_1 s_2 p_3 +s_1 s_3 p_2 + s_2 s_3 p_1) \sin{(\varphi_1+\varphi_2+\varphi_3)} \Biggr]. \nn
\ee

\end{widetext}

Importantly, the first line here constitutes the free energy of the 1$Q$ phase if one puts $s_\Sigma=s, \, p_\Sigma=p$ (cf. Eq.~\eqref{F1Q}), whereas the second and the third lines determine the ``penalty''  and the ``profit'' for having 3$Q$ structure instead of 1$Q$, respectively. They can be considered as effective ``cubic anisotropy'' in the order parameter space. Evidently, the superposition of two helicoids has no advantages in this model, because the last term in Eq.~\eqref{AF3Q1} is zero. So, we left with two possibilities: one can have a single helicoid component (e.g., with nonzero $s_1$ and/or $p_1$), or all three helicoids. In the former case we arrive to the free energy given by Eq.~\eqref{F1Q}, while in the latter case it can be checked, that the minimum of the free energy corresponds to the spin structure~\eqref{S3Q1} with all equal $s_i$ and $p_i$, and $\sum_i \varphi_i =  \pi/2 + \pi n$ with integer $n$ (the signs of $s$ and $p$ should be properly chosen). This leads to the free energy of the 3$Q$ phase in the form of Eq.~\eqref{F3Q}.

Analytical minimization of the free energy~\eqref{F3Q} (using the trick with magnetization described above) leads to a system of cubic equations:
\be \label{AEqS}
 &&-t_h s + b \left( \frac{3}{2}s^3 + \frac{5}{6} s p^2 - \frac{2 \chi h s p}{\sqrt{3}}\right) = 0, \\
&&-(t_h-\Lambda_h) p + b \left( \frac{5}{2}p^3 + \frac{5}{6} s^2 p - \frac{\chi h (s^2 + 6 p^2)}{\sqrt{3}}\right) = 0. \nn
\ee
The first one gives either $s=0$ or
\be \label{ASolS}
  s^2= \frac{2}{3} \left(\frac{t_h}{b} - \frac{5}{6}p^2 + \frac{2 \chi h p}{\sqrt{3}} \right),
\ee
which can be plugged into the second equation in the system~\eqref{AEqS}. After some algebra, we arrive to the following cubic equation for $p$:
\be \label{ACubEq}
  &&p^3 - \frac{39 \sqrt{3} \chi h}{55}p^2 + \left[ \frac{24}{11}(\chi h)^2+ \frac{27}{55b}\left( \Lambda - \frac{4}{9} t \right)\right]p \nn \\
  &&-\frac{6\sqrt{3} \chi h t_h}{55b}=0.
\ee
This equation can be solved using Cardano's formula. To make the corresponding result more compact, first we introduce
\be
  \alpha &=& \frac{39 \sqrt{3} \chi h}{55}, \nn \\
  \beta &=&  \frac{24}{11}(\chi h)^2+ \frac{27}{55b}\left( \Lambda - \frac{4}{9} t \right),  \nn \\ \label{ACubPar}
  \gamma &=& \frac{6\sqrt{3} \chi h t_h}{55b}, \\
  \rho &=& \alpha^2 - 3 \beta, \nn \\
  \sigma &=& 2 \alpha^3 - 9 \alpha \beta + 27 \gamma. \nn
\ee
Using these expressions we can write solutions for $p$ in the form:
\be \label{ACubSol}
  p &=& \frac{\alpha}{3} - \frac{(-1)^{1/3}2^{1/3} \rho}{3 (\sigma + \sqrt{\sigma^2 -4 \rho^3})^{1/3}} \nn \\ &&- \frac{(\sigma + \sqrt{\sigma^2 -4 \rho^3})^{1/3}}{(-1)^{1/3}2^{1/3}3},
\ee
where $(-1)^{1/3} = -1, (1+i\sqrt{3})/2, (1-i\sqrt{3})/2$; these values should be plugged simultaneously in both the second and the third terms of Eq.~\eqref{ACubSol}, whereas the other cube roots should be taken for the branch $(-1)^{1/3} = -1 $.

The 3$Q$ phase is (meta)stable at $t > 2 b(\chi h)^2$, where there is an instability towards nonzero $s$ value, which, in its turn, leads to nonzero $p$ in external field due to the term $\propto m p s^2$ (see Eq.~\eqref{F3Q}). Analysis shows, that in this region of the phase diagram, the proper solution for $p$ is given by Eq.~\eqref{ACubSol} with $(-1)^{1/3}=-1$. Then, one can calculate $s$ using Eq.~\eqref{ASolS}. Finally, plugging these particular $p$ and $s$ into Eq.~\eqref{F3Q} with $m = \chi h$, one could obtain analytical expression for 3$Q$ phase free energy. We will not write down it here explicitly, because it is cumbersome.

\subsection{Phase boundaries}

\begin{figure}
  \centering
  \includegraphics[width=8cm]{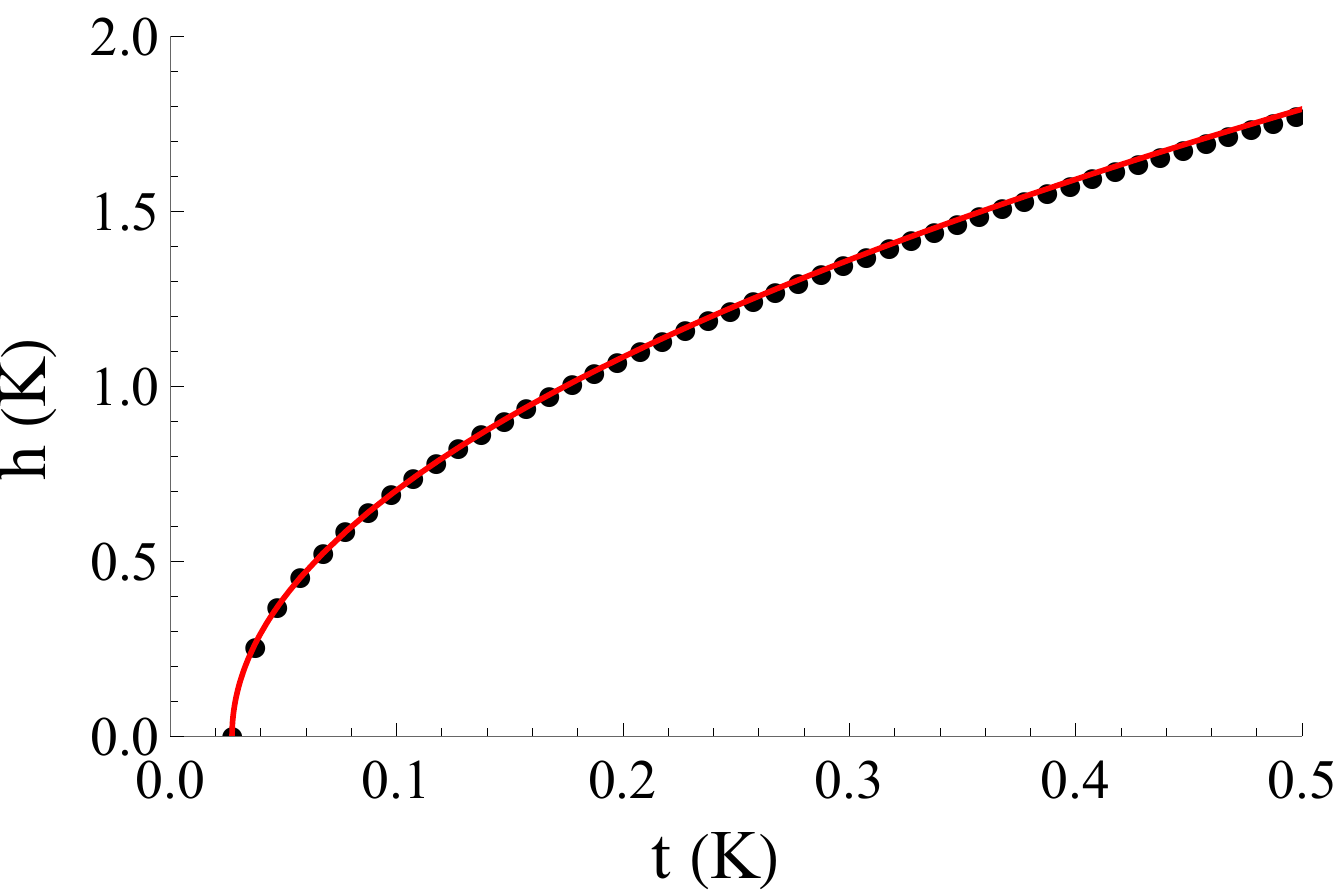}\\
  \caption{Illustration for empirical law~\eqref{At1Q3Q1} for the boundary between 1Q and 3Q phases. Black dots stand for numerical solution of $\mathcal{F}_{1Q} = \mathcal{F}_{3Q}$ equation, whereas the red line is for Eq.~\eqref{At1Q3Q1}.  Parameters set~\eqref{param1} is used. }\label{F1Q3QApp}
\end{figure}

\begin{figure}
  \centering
  \includegraphics[width=8cm]{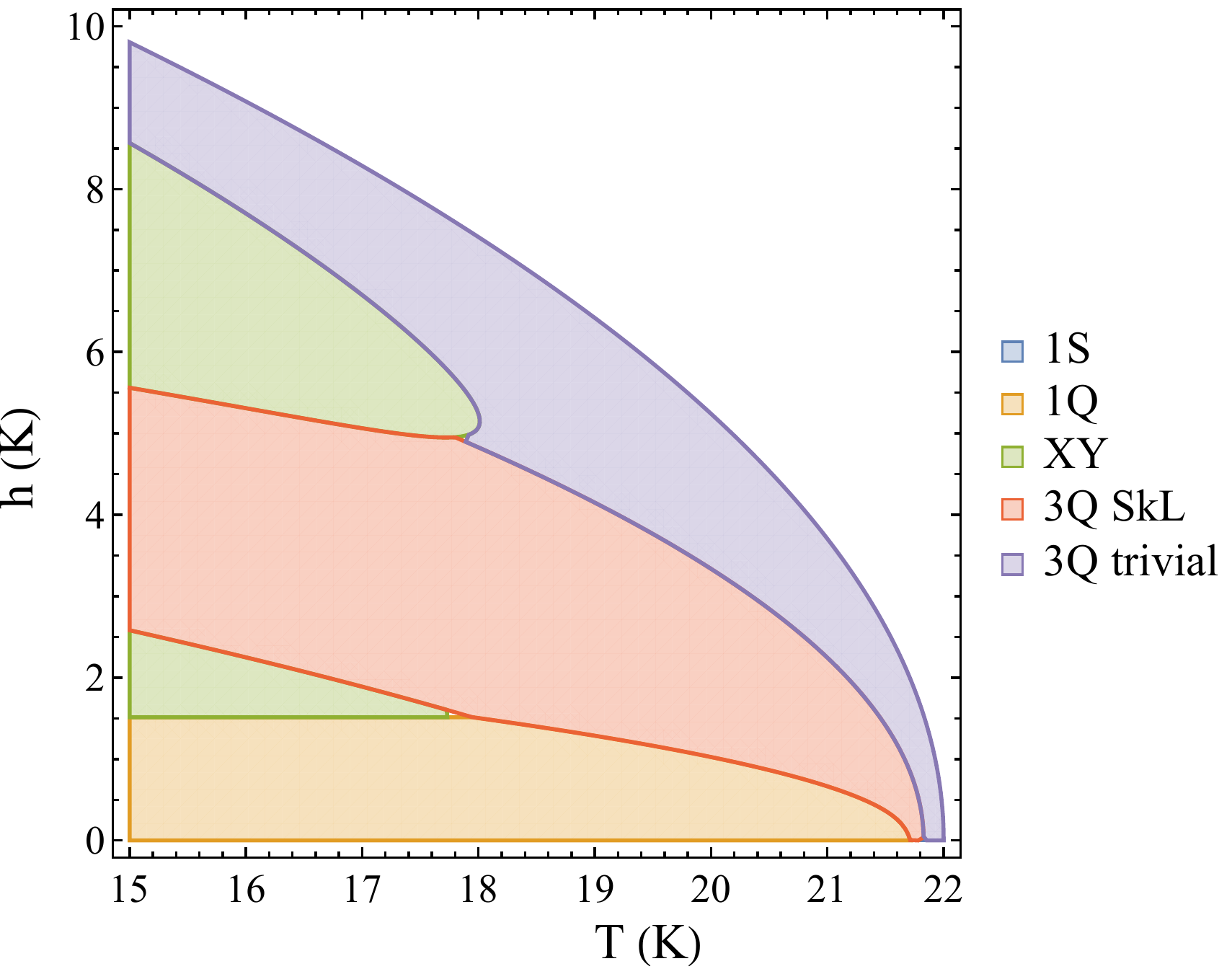}\\
  \caption{If one puts parameter $\Lambda^\prime$ in the set~\eqref{param1} to be smaller, then the XY phase appears on the phase diagram (cf. Fig.~\ref{FPhase1}). However, the 3Q SkL phase is still the ground state in the large part of the phase diagram. }
  \label{FPhaseXY1}
\end{figure}

After free energy derivation, we can turn to analysis of the boundary between 1$Q$ and 3$Q$ phases. Approximate expression for this curve was obtained by fitting numerical data for various parameters sets. The result is the following:
\be \label{At1Q3Q1}
  t_{1Q-3Q}(h) \approx \frac{3}{2} \Lambda + 45 b (\chi h)^2.
\ee
This expression works quite well for not very small $h$, where linear in $h$ term is somewhat important. Particular usage of this expression for the parameters set~\eqref{param1} is shown in Fig.~\ref{F1Q3QApp}. Note, that at given $h$ Eq.~\eqref{At1Q3Q1} yields much larger temperature, than Eq.~\eqref{At1Q}. For instance, at $h_{SF}$ (see Eq.~\eqref{AhSF1}) one has $t_{1Q-3Q} \approx 3 \Lambda/2 + 11.25 (\Lambda^\prime- \Lambda)$, which is a very large quantity for our approach (note, that for the parameters set~\eqref{param1} it corresponds to $T<0$). Nevertheless, it shows that the 3$Q$ phase should be stable even at temperatures substantially lower than the ordering one ($T_c$). This can be also illustrated by manually making $\Lambda^\prime$ in the parameters set~\eqref{param1} to be much smaller than $0.26$~K, e.g., equal to $0.05$~K. Then, the XY phase appears at larger temperatures, and its traces become visible in the phase diagram, see Fig.~\ref{FPhaseXY1}. 

Topological properties of the 3$Q$ phase depend on whether the spins can wrap around the whole sphere in the spin space or not. In the external field along $\m{c}$ axis, it requires that the maximal negative value of modulated $z$-component of spin (which is equal to $-3 p/\sqrt{3}$) overcomes positive magnetization $m$. So, the boundary between 3Q trivial and 3Q SkL phases is given by equation $p=m/\sqrt{3}$. We plug it into Eq.~\eqref{ACubEq}. Then, it can be shown that this equation is satisfied if
\be
  t = t_{SkL} &=& \frac{9}{10} \Lambda + \frac{203}{45} b (\chi h)^2 \nn \\ &\approx& 0.9 \Lambda + 4.5 b (\chi h)^2.
\ee
At larger $t$ (smaller temperatures $T$) the structure is topologically nontrivial with $n_{Sk} = \pm 1$ (it depends on the choice of three constituent helicoids chiralities) per magnetic unit cell.

\section{Free energies of various phases and phase diagram for out-of-plane collinear easy axes}
\label{AppendB}

When the easy axes for possible modulation vectors $\m{k}_j$ are collinear and oriented along the $\m{c}$ axis, the formulas are similar with those of Appendix~\ref{AppendA}, however some important differences appear. Here we discuss them in details and obtain the corresponding phase boundaries.

In the external magnetic field it is convenient to introduce the following renormalized parameters (cf.~\eqref{th} and~\eqref{Lh}), which are used in equations below:
\be \label{Ath1}
  t_h = t - 6 b (\chi h)^2, \\ \label{ALh1}
  \Lambda_h = \Lambda - 4 b (\chi h)^2.
\ee

\subsection{Simple spin-density wave (1$S$ phase)}

Here the spin polarization is along the easy axis $\m{c}$, so it is parallel to magnetic field:
\begin{equation}\label{AS1S2}
  \mathbf{s}_i = (s\cos{\mathbf{k}_1 \mathbf{R}_i} + m )\hat{z}.
\end{equation}
For $s$ and $\mathcal{F}_{1S}$ one should use Eqs.~\eqref{AS1S1} and~\eqref{AF1S1}, but with $t_h$ defined in Eq.~\eqref{Ath1}. The boundary between PM and 1$S$ is thus defined by condition
\be
  t_h = 0 \Leftrightarrow t = 6 b (\chi h)^2.
\ee

\subsection{Single-Q elliptical helicoid (1$Q$ phase)}

We left the definition of this spin ordering~\eqref{S1Q1} intact, so in comparison with Eqs.~\eqref{AS1Q1} the spin components should be interchanged:
\be \label{AS1Q2}
  p^2 &=& \frac{2 t_h + \Lambda_h}{4b}, \nn \\
  s^2 &=&  \frac{2 t_h - 3 \Lambda_h}{4b},
\ee
where conditions $t_h > 3 \Lambda_h/2$ and $t_h > - \Lambda_h/2$ should hold (the last one can be violated in the external field). Importantly, the former one defines the boundary between the 1$S$ and 1$Q$ phases, which is a simple vertical line:
\be
  t_h = 3 \Lambda_h/2 \Leftrightarrow t = 3 \Lambda/2.
\ee
The free energy of 1$Q$ is given exactly by Eq.~\eqref{AF1Q1}.

\subsection{Conical helicoid (XY phase)}

The modulated components of spins in the conical phase are rotating in the hard plane. As compared to the previous 1$Q$ case, substitutions $t_h \rightarrow t^\prime_h - \Lambda$ and $\Lambda_h \rightarrow \Lambda^\prime- \Lambda $ are in order, where $t^\prime_h = t - 2 b (\chi h)^2$. So, XY phase free energy reads
\be \label{AFXY2}
  \mathcal{F}_{XY} &=& - \frac{4 (t^\prime_h - \Lambda)^2 - 4 (t^\prime_h - \Lambda) (\Lambda^\prime - \Lambda) + 3 (\Lambda^\prime - \Lambda)^2}{16 b} \nn \\ &&- \frac{\chi h^2}{2}, \quad t^\prime_h > \frac{3\Lambda^\prime - \Lambda}{2}.
\ee
Importantly, the spiral plane flop field in this case reads
\be \label{AhSF2}
 h_{SF} = \sqrt{\frac{\Lambda^\prime}{4 b \chi^2}},
\ee
which is larger than the one for the in-pane easy axes case (see Eq.~\eqref{AhSF1}).

\subsection{Superposition of three screw helicoids (3$Q$ phase)}

Here one should make the following substitution in the free energy~\eqref{AF3Q1}:
\be
  -\frac{t}{2}s^2_\Sigma - \frac{t-\Lambda}{2}p^2_\Sigma \rightarrow -\frac{t}{2}p^2_\Sigma - \frac{t-\Lambda}{2}s^2_\Sigma,
\ee
which leads to qualitatively new behaviour of order parameters in comparison with the case of in-plane easy axes.

The counterparts of Eqs.~\eqref{AEqS} read
\be \label{AEqS2}
 &&-(t_h-\Lambda_h) s + b \left( \frac{3}{2}s^3 + \frac{5}{6} s p^2 - \frac{2 \chi h s p}{\sqrt{3}}\right) = 0, \nn \\
&&-t_h p + b \left( \frac{5}{2}p^3 + \frac{5}{6} s^2 p - \frac{\chi h (s^2 + 6 p^2)}{\sqrt{3}}\right) = 0.
\ee
We can rewrite the first equation as
\be \label{ASolS2}
  s  \left[\frac{3}{2} b s^2  + \frac{5b}{6}p^2 - \frac{2 b \chi h p}{\sqrt{3}} - (t_h - \Lambda_h) \right] = 0.
\ee

Assuming $s=0$ in the second equation of the system~\eqref{AEqS2}, and plugging its $p$-dependent part into equation~\eqref{ASolS2}, we find that there are no additional solutions for $s$ if $t_h \leq 3 \Lambda_h/2 \Leftrightarrow t \leq 3 \Lambda/2$. Then, we arrive to an interesting conclusion: in this region of parameters the 3$Q$ phase is just a superposition of three collinear SDWs. We will refer to this spin structure as 3$P$. It is illustrated in Fig.~\ref{Fig3P}. Evidently, this phase is topologically-trivial. Corresponding order parameter $p$ can be simply obtained from the quadratic equation; the result being
\be \label{AS3P1}
  p = \frac{1}{5} \left( 2 \sqrt{3} \chi h  + \sqrt{12 (\chi h)^2 + 10 t_h/b}\right).
\ee
Together with $s=0$ and $m=\chi h$ it allows to calculate 3$P$ phase free energy.

For $t> 3 \Lambda/2$ parameter $s$ is nonzero, and (as in Appendix~\ref{AppendA}) one needs to solve cubic equation:
\be \label{ACubEq2}
  &&p^3 - \frac{39 \sqrt{3} \chi h}{55}p^2 + \left[ \frac{24}{11}(\chi h)^2 - \frac{3}{55b}\left( 4 t + 5 \Lambda \right)\right]p \nn \\
  &&-\frac{6\sqrt{3} \chi h (t_h-\Lambda_h)}{55b}=0.
\ee
In order to utilize solutions~\eqref{ACubSol} one should use modified parameters
\be
  \beta &=& \frac{24}{11}(\chi h)^2 - \frac{3}{55b}\left( 4 t + 5 \Lambda \right), \\
  \gamma &=& \frac{6\sqrt{3} \chi h (t_h-\Lambda_h)}{55b}, \nn
\ee
along with $\alpha$, $\sigma$ and $\rho$ from Eqs.~\eqref{ACubPar}.

\subsection{Phase boundaries}

We start from the phase boundaries at $t \leq 3 \Lambda/2$. In this region, the competing phases are PM, 1$S$ and 3$P$. By comparing free energies, we find that there is a boundary between 1$S$ and 3$P$ which reads
\be \label{At1S3P}
  t_{1S-3P} &=& \left[ \frac{\sqrt{6}-1}{(\sqrt{6}-2)^2} \frac{24}{5} +6 \right] b (\chi h)^2 \\
  &\approx& 40 b (\chi h)^2. \nn
\ee
At given $t$ the 1$S$ phase is stable below this curve (at smaller fields) in the $T-H$ plane, and 3$P$ is stable above it.

At yet higher magnetic fields, there is a first order transition between 3$P$ and PM, which is evident from the 3$P$ phase free energy (bearing in mind that $s=0$ and $m=\chi h$), which has the form (cf. Eq.~\eqref{F3Q}):
\be \label{AF3P}
  \mathcal{F}_{3P}= -\frac{t_h}{2}p^2- \frac{\chi}{2} h^2 +
    b \Biggl[ \frac{5 p^4}{8} - \frac{2 \chi h p^3}{\sqrt{3}} \Biggr].
\ee
The cubic term here induces discontinuous transition. After some algebra we find the phase boundary
\be \label{At3PPM}
  t_{3P-PM} &=& \frac{74}{15} b (\chi h)^2.
\ee
Note, that $t_{3P-PM}$ is always smaller, than the $t_{1S-3P}$ at given $h$, so there is no boundary between 1$S$ and PM phases.

At $t > 3 \Lambda/2$ nonzero $s$ appears. So, 1$Q$, 3$Q$ and PM are competing. The boundary between 1$Q$ and 3$Q$ is hard to found explicitly, however, we observe that Eq.~\eqref{At1S3P} describes this curve quite accurately. In the high fields domain the transition between 3$Q$ SkL and PM can be either direct (first order one) or via the intermediate 3$Q$ trivial phase. This should be contrasted with the in-plane easy axes case where 3$Q$ trivial always appears before PM. The 3$Q$ trivial phase continuously transforms into PM at
\be \label{At3QPM}
  t_h=\Lambda_h \leftrightarrow t = \Lambda + 2 b (\chi h)^2.
\ee
Moreover, the boundary (if exists) between 3$Q$ SkL and 3$Q$ trivial can be found analytically from the condition $p = m/\sqrt{3}$. Plugging the latter into the cubic equation~\eqref{ACubEq2} we can found the corresponding curve
\be \label{AtSkL2}
  t_{SkL} = \frac{\Lambda}{10} + \frac{203}{45} b (\chi h)^2.
\ee
Importantly, the curves~\eqref{At3QPM} and~\eqref{AtSkL2} intersect at $t^\prime = 194 \Lambda/113 \approx 1.72 \Lambda$ which is larger than $3 \Lambda/2$. It means that below $t^\prime$ the 3$Q$ trivial phase does not exist. In the range $t \in (1.5 \Lambda, 1.72 \Lambda) $ there is a first order transition between 3$Q$ SkL and PM phases. The boundary between them interpolates two curves~\eqref{At3PPM} and~\eqref{AtSkL2}. At $t>t^\prime$ the line of the first order transitions (here between 3$Q$ SkL and 3$Q$ trivial) is approximately given by Eq.~\eqref{AtSkL2}. It terminates at $t=t^{\prime \prime}$ (it will be quantified below). In the region of temperatures $t^\prime<t<t^{\prime\prime}$ at relevant magnetic fields, $t_{SkL}$ corresponds to spurious real solution for $p$, whereas the physical solution discontinuously jumps from $p>m/\sqrt{3}$ to $p<m/\sqrt{3}$. At $t>t^{\prime\prime}$ there is only one physical solution, which continuously changes with $h$ and there is a smooth crossover between 3$Q$ SkL and 3$Q$ trivial phases at $t_{SkL}$ given by Eq.~\eqref{AtSkL2}.

Finally, we discuss analytical derivation of $t^{\prime\prime}$. At this temperature the plot for $p(h)$ has vertical tangent; at $t<t^{\prime\prime}$ there is a region of fields with three real solutions and at $t>t^{\prime\prime}$ there is only one real solution in the high-fields domain (near the PM phase stability part of the phase diagram). In order to obtain $t^{\prime\prime}$ we rewrite Eq.~\eqref{ACubEq2} in the form
\be \label{ACubEq3}
  &&p^3 - \frac{39 \sqrt{3} \chi h}{55}p^2 + \left[ \frac{24}{11}(\chi h)^2 - \frac{3}{55b}\left( 4 t + 5 \Lambda \right)\right]p \nn \\
  &&=\frac{6\sqrt{3} \chi h (t_h-\Lambda_h)}{55b}.
\ee
where the r.h.s. is $p$-independent. Then, the solutions can be found graphically, as intersections of cubic parabola and the horizontal line. At $t^{\prime\prime}$ the l.h.s. should have horizontal tangent and inflection at the same point. So, conditions $\partial \text{l.h.s.}/\partial p =0$ and $\partial^2 \text{l.h.s.}/\partial p^2 =0$ should hold simultaneously with Eq.~\eqref{ACubEq3}. The latter simply yields $p = 13 \sqrt{3} \chi h/ 55$; plugging it into the former we get $4 t^{\prime\prime} + 5 \Lambda = 1693 b (\chi h)^2/55 $. Next, using this two formulas and Eq.~\eqref{ACubEq3} we obtain
\be
  t^{\prime\prime} = \frac{2\times 93115 - 5 \times 14297}{2\times 93115 + 4 \times 14297}\Lambda \approx 0.2 \Lambda.
\ee
Note, that $t^{\prime\prime}>t^\prime>1.5\Lambda$, so discussed above topology of the phase diagram should be independent on parameters of a model, until the easy axes are collinear.

\bibliography{TAFbib}

\end{document}